\documentclass[runningheads,a4paper]{llncs}

\usepackage{amssymb}
\usepackage{graphicx}
\usepackage{url}

\usepackage{listings}
\usepackage{multirow}
\usepackage{algorithm}
\usepackage{algorithmic}
\usepackage{subfigure}
\usepackage{amsmath}
\usepackage{cite}

\urldef{\mailsa}\path|{lei.gai, pekingchenwei, zhchxu, chhqiu, tjwang}@pku.edu.cn|

\begin{document}

\mainmatter  

\title{Towards Efficient Path Query on Social Network with Hybrid RDF Management \thanks{This research is supported by the National High Technology Research and Development Program of China (Grant No. 2012AA011002), Natural Science Foundation of China (Grant No. 61300003), Specialized Research Fund for the Doctoral Program of Higher Education(Grant No. 20130001120001).}}

\titlerunning{Towards Efficient Path Query with Hybrid RDF Management}

%
%
\author{Lei Gai \and Wei Chen \thanks{Corresponding author.}\and Zhichao Xu\and Changhe Qiu\and Tengjiao Wang
}
\authorrunning{Lei Gai et al.}

\institute{School of Electronic Engineering and Computer Science, \\ Peking University, Beijing, China \\
\mailsa}
%
%
\toctitle{Lecture Notes in Computer Science}
\tocauthor{Lei Gai et al.}
\maketitle

\thispagestyle{empty}
\pagestyle{empty}
%
%

\begin{abstract}
The scalability and flexibility of Resource Description Framework(RDF) model make it ideally suited for representing Online Social Networks(OSN). One basic operation in OSN is to find chains of relations, such as $k$-Hop friends. Property path query in SPARQL can express this type of operation, but its implementation suffers from performance problem considering the ever growing data size and complexity of OSN.

In this paper, we present a main memory/disk based hybrid RDF data management framework for efficient property path query. This hybrid framework realizes an efficient in-memory algebra operator for property path query using graph traversal, and estimates the cost of this operator to cooperate with existing cost-based optimization. Experiments on benchmark and real dataset demonstrated that our approach achieves a good tradeoff between data load expense and online query performance.
\end{abstract}
\section{Introduction}
In the age of Web 2.0, OSN have gained pervasive interests in both research communities and industries. There is a trend to model OSN using Semantic Web technologies, especially the vocabularies from
FOAF\footnote{\url{http://www.foaf-project.org/}} and SIOC\footnote{\url{http://rdfs.org/sioc/spec/}} project. RDF, originally designed for the Semantic Web, have been wildly adopted for representing such kind of linked data. SPARQL\footnote{\url{http://www.w3.org/TR/rdf-sparql-query/}} as the de-facto RDF query language is used for Basic Graph Pattern(BGP) query with bounded or unbounded variables. The scalable graph representation and flexible query capability make RDF model suited for large-scale complex OSN management and analysis.
\begin{figure}
    \centering
    \includegraphics[width=3 in]{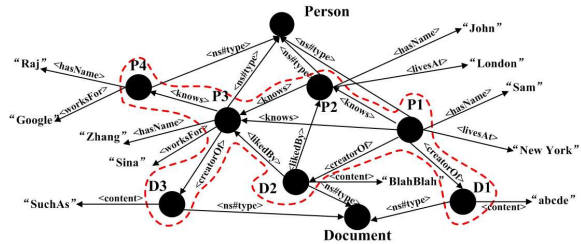}
    \caption{Examples for a fraction of social network.  }
    \label{fig:graph-pattern}
\end{figure}

Figure \ref{fig:graph-pattern} illustrated a snippet of large social graph representing relations between four users and three User Generated Contents(UGC).  Query which \textit{Find pair of users in a path of friend relationship which user2 has a job and like the documents created by user1 } is expressed in SPARQL as:
\begin{lstlisting}[
           language=SQL,
           showspaces=false,
           basicstyle=\ttfamily,
           commentstyle=\color{gray}
        ]
SELECT DISTINCT ?user1, ?user2 WHERE {
    ?user1 knows* ?user2 .
    ?user1 creatorOf ?doc1 .
    ?user2 worksFor ?organization .
    ?doc1 likedBy ?user2 }
\end{lstlisting}
For RDF graph in Figure \ref{fig:graph-pattern}, this query returns the result set $R(q) =\{<P1, P3>\}$. The pattern \emph{?user1 knows* ?user2} states the path consist of zero or more \emph{knows} predicates, it is a property path query pattern.

\vspace{6pt} \noindent \textbf{Challenges.} Path queries are of common interest in OSN analysis for discover complex relations among entities. Despite the scalability and flexibility provided by RDF model, Path queries performs poorly and lack of efficient implementation in existing RDF management. Current researches in graph analysis domain are mainly based on pre-constructed reachability indices. Building such indices are both time and memory consuming, especially when dealing with large complex graph. From RDF management point of view, due to costly join for triple pattern matching, there lacks efficient implementation of property path query. Although property path query is recommended by newest SPARQL 1.1 standard, to the best of our knowledge Jena\footnote{\url{https://jena.apache.org/}}, Virtuoso\footnote{\url{http://virtuoso.openlinksw.com/}} and Sesame\footnote{\url{http://www.openrdf.org/}} are the only three off-the-shelf RDF stores that support standardized path query. They all suffer from performance problems when dealing with path query on large-scale data.

\vspace{6pt} \noindent \textbf{Overview of Our approach.} Our approach is motivated by three observations. First, current
off-the-shelf RDF store performs well only on join-based star queries. Second, the search space of property path query is restricted to triples representing relations among entities. Third, due to the rich semantics in OSN, most triples are for entities attributes not for relations. Based on these observations, we argue that manage graph topology-related triples into main memory is feasible, and this will greatly enhance the online query performance of property path related sophisticate SPARQL queries. We propose a hybrid framework that has in-memory management of graph topology-related RDF data along with disk-based Jena TDB\footnote{\url{https://svn.apache.org/repos/asf/jena/}} native triple store. In our approach, while loading and indexing triples into TDB, graph topology-related triples are identified and duplicated in main memory. For an online query, we implement a graph traversal based algebra operators for property path pattern, which is more efficient than traditional join-based operator.

\vspace{6pt} \noindent \textbf{Contributions.} We summarized our contribution in this paper as:
\begin{enumerate}
    \item We propose a main memory/disk based hybrid framework for efficient property path query. While leverage the functionality of existing  well-established RDF store for BGP query, it specialized in property path pattern query through manage graph topology  related data in main memory.
    \item We present an algebra operator for property path pattern query. Its realization based on in-memory graph traversal instead of costly join. Using the characteristics of OSN, heuristics for estimating the execution cost of this operator is given that can be used for cost-base optimization.
\end{enumerate}

\vspace{6pt} \noindent \textbf{Organization of the paper.} The rest of the paper is organized as follows. Section \ref{sec:hybrid} presents the basic design of
our hybrid RDF management framework. Section \ref{sec:evaluation} shows the evaluation results. We conclude in Section \ref{sec:conclusion}.
\section{Hybrid RDF Data Management}\label{sec:hybrid}
Given vocabulary $\Sigma = V_{E} \cup V_{A} \cup E_{EE}  \cup E_{EA} \cup L_{E} \cup L_{A} $ defined in Table \ref{table:notations}, an OSN is represented as a triple set $T_{OSN} = T_G \cup T_A $, where the graph-topology set $T_G \subseteq V_{E} \times L_{E} \times V_{E}$ holds triples representing social entities and relations among them, and the attributes set $T_A \subseteq V_{E} \times L_{A} \times V_{A}$ holds triples representing attributes and theirs relations to social entities.
\begin{table}[ht]
\label{table:notations}
\caption{Notations for social graph representation.}  
\centering   
\begin{tabular}{ | c | c | c |}   
\hline    
Notations & Refers to the set of & Instance of Figure 1 \\[0.5ex]  
\hline    
$V_{E}$ & Nodes for social entities.  & $\{P1,D1,\ldots\}$  \\  
$V_{A}$ & Nodes for attributes values. & $\{"John", "London", "abcde", \ldots\}$    \\
$E_{EE}$ & Edges among $V_{E}$. & $\{(P1,P2),(P1,D1), \ldots\}$    \\
$E_{EA}$ & Edges between $V_{E}$ and $V_{A}$.  &  $\{(P1,"Sam"),(D1,"abcde"), \ldots\}$   \\
$L_{E}$ & Labels for $E_{EE}$. & $\{<knows>,<likeBy>,\ldots\}$   \\
$L_{A}$ & Labels for $E_{EA}$. & $\{<hasName>, <ns\#type>, \ldots\}$   \\[1ex] 
\hline    
\end{tabular}
\label{table:nonlin}
\end{table}
\subsection{Architectural Design}\label{subsec:architecture}
\begin{figure}
    \centering
    \includegraphics[width=3.2in]{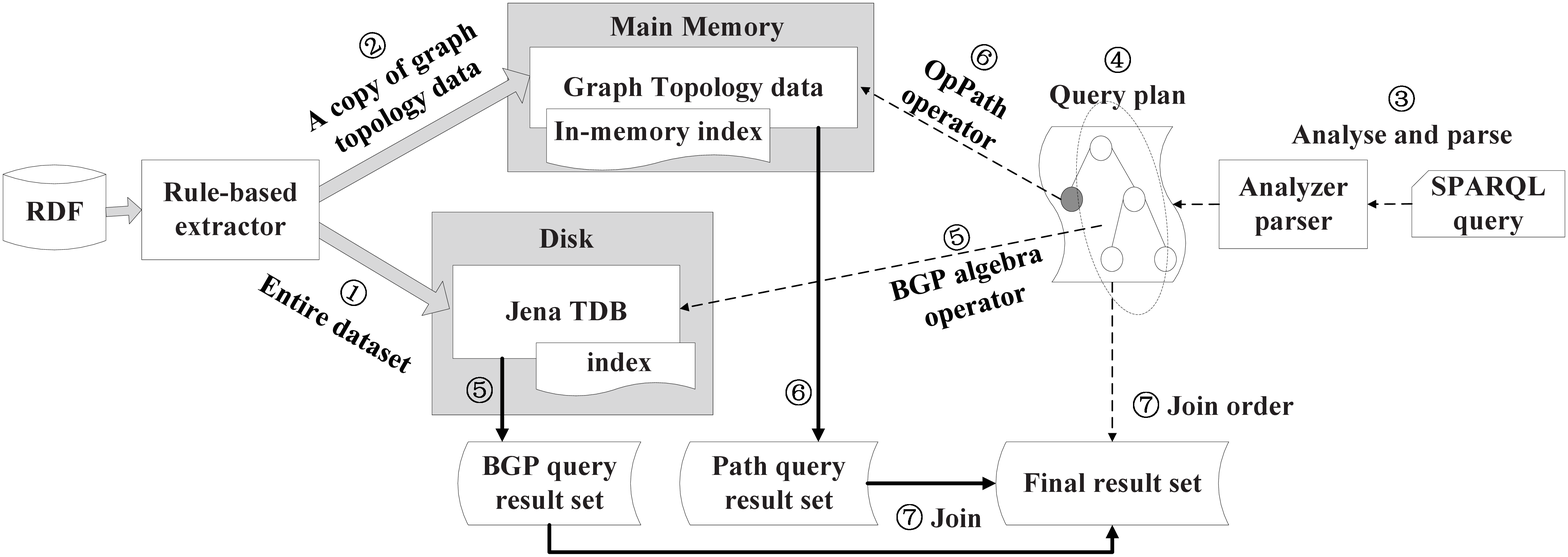}
    \caption{Hybird RDF management architecture of our approach }
    \label{fig:architecture}
\end{figure}

In this paper, we focus on efficient implementation of property path query. Consider that path query only related to Triples in $T_G$ (In Figure \ref{fig:graph-pattern}, $T_G$ is represented as dashline-encircled part), one direct motivation is that manage $T_G$ in memory will greatly enhance the overall query performance. Based on this motivation, we implement a hybrid RDF data management architecture that manage different query-prone data respectively. This architecture is shown in Figure \ref{fig:architecture}.

Our approach can be thought as a plugin component which override the functionality of corresponding part in TDB. At data loading stage,  $\forall t_i \in T_{OSN}$ is loaded into Jena TDB (step \textcircled{1} in Figure \ref{fig:architecture}). At the same time, $\exists t_i \in T_G$ is filtered out based on two kinds of rules:
\begin{enumerate}
  \item The type of \textit{Objects}. If \textit{Objects} for $t_i$ is a literal, then $t_i \in T_A$ .
  \item The semantic meaning of \textit{Predicate}. Such as \textit{foaf:knows} defined in FOAF project which state the relation  that \emph{Subject User} know \emph{Object User}.
\end{enumerate}

$T_G$ is stored in main memory with subject index(PSO) for forward traversal and object index(POS)for backward traversal(step \textcircled{2}).These indices can be constructed incrementally when topology-related data is extracted and loaded into main memory.

When an online query is submitted, SPARQL parser translates query strings into patterns based on standard abstract syntax tree recommended by W3C (step \textcircled{3}). Analyzer translates these patterns into predefined SPARQL algebra operators and construct execution plan (step \textcircled{4}).An algebra operator generates designated result set from given input (step \textcircled{5}).  In our approach we implemented a special operator named \textit{OpPath} which only uses in-memory data as input. If a query string in \textit{WHERE} clause is analyzed as property path pattern, \textit{OpPath} operator is added to the query plan.  (step \textcircled{6}) We explain the design of \textit{OpPath} operator in  detail in Section \ref{subsec:operator}. Result set of algebra operator is joined to get the final result set. The join order of operators is optimized using cost and selectivity estimation (step \textcircled{7}).
\subsection{Property Path Algebra Operator}\label{subsec:operator}
\begin{definition}[\textit{OpPath} Operator]
\textit{OpPath} is a ternary algebra operator that can be defined as \textit{OpPath(O,S,$P_P$)}. $S,O \subseteq V_{EE}$ can be either bounded or unbounded variables, $|S| =s$, $|O| =o$ . $P_P$ is a regular pattern expression defining the property path. \textit{OpPath(O,S,$P_P$)} operator find existing path from set $S$ to set $O$, and return all triple sets that each paths is consist of as result set.
\end{definition}
Based on research\cite{zeng2013distributed} which have testified that graph explorations is extremely efficient and more easy to implement than costly joins, the \textit{OpPath} operator is realized as in-memory Breadth-First Search(BFS). The \textit{OpPath} operator has time complexity $O(|V_E|+|E_{EE}|)$ and space complexity $O(|T_G|)$. It is much less than tradition nested-loop join that has time complexity $O(|V_E| \cdot |E_{EE}|)$.

The cost of \textit{OpPath} operator is the cardinality of result set $R(q)$ for path query pattern $q$. Existing researches such as G-SPARQL\cite{sakr2012g} using predefined heuristics which always take $|R(q)|$ as the largest, this is far from optimal. Sparqling Kleene\cite{gubichev2013sparqling} using pre-computed reachability path indices which affects data load efficiency. We consider three factors that affects $|R(q)|$, the average nodes out-degree, the path length $l$ and the pathes that fits for the given pattern. In our approach, we leverages the graph generation model \cite{leskovec2007graph} which expects the average out-degree as $d_{out}=|V_{EE}|^{1-\ln c}, 1<c\leq 2$. We also assumes that nodes in $T_G$ has the same probability of being added to the path, thus the modifying factor of out-degree follows the binomial distribution. For all considerations above, $|R(q)|$ can be approximately estimated as:
\begin{equation}
\label{equ:cost}
|R_q| = s\cdot o \cdot \sum_{i=1}^{l}(|V_{EE}|^{(1-\ln c) \cdot i} \cdot (\sum_{i=1}^{l}\frac{l!}{i!(l-i)!} \cdot p^i \cdot(1-p)^{l-i} ))
\end{equation}
where $p=\frac{|E_{EE}|-|V_{EE}|}{|V_{EE}|}$. $|V_{EE}|$ and $|E_{EE}|$ can be got from the metadata in RDF store. We performs preliminary testing to measure the accuracy of Equation \ref{equ:cost} with real all-pair cardinality of dataset in Table \ref{table:datasets}. For SNIB $T_G$ with average $d_{out}=12$, $c=1.75$, $relative\ error = \frac{max(real\ cardinality ,estimate\ cardinality)}{min(real\ cardinality ,estimate\ cardinality)}-1$ is about $27\%$. For DBLP $T_G$ with $d_{out}=7$, $c=1.81$, and $error = 32\%$. This preliminary testing shows that the heuristic defined in Equation \ref{equ:cost} is with acceptable cardinality estimation error.
\section{Evaluation}\label{sec:evaluation}
We used one machine with Debian 7.4 in 64-bit Linux kernel, two Intel Xeon E5-2640 2.0GHz processor and 64 GB RAM for our evaluation. Our approach is compared with Jena(version 2.11.1), Sesame(version 2.7.10) and competitive research G-SPARQL\cite{stocker2008sparql}. Jena implements path query based on join while Sesame is based on graph traversal. We also implements our approach with no cost estimation and treats path query as the most costly(denoted as \textit{NoCE}). All evaluations were done 10 times and the results are the averages.

We adopted two datasets, the Social Network Intelligence Benchmark (SNIB)\footnote{Social Network Intelligence Benchmark(SNIB), \url{http://www.w3.org/wiki/Social_Network_Intelligence_BenchMark/}.} as synthetic dataset, and DBLP as real dataset, all in RDF N-Triples format\footnote{\url{http://www.w3.org/TR/n-triples/}}. SNIB dataset is generated using S3G2\cite{pham2013s3g2}. Considering G-SPARQL uses ACM digital library dataset which is not publicly available, we uses the DBLP dataset instead \footnote{DBLP dataset can be download from \url{http://sw.deri.org/~aharth/2004/07/dblp/dblp-2006-02-06.rdf}. Same as stated in G-SPARQL, we manually created the co-author relationships between author nodes, which originally recorded as $<creator>$ tag in raw dataset.}. Statistics of datasets are shown in Table \ref{table:datasets}. The DBLP dataset has approximately the same characteristics as the \textit{Large Graph Size} experiment in G-SPARQL.
\begin{table}[ht]
\caption{Statistics for SNIB and DBLP datasets.}  
\label{table:datasets}
\centering   
\begin{tabular}{ | c | c | c |  c  | c |}
\hline    
Dataset & Vertices($|V_{EE}|$) & Edges ($|E_{EE}|$) & Attributes ($|T_{A}|$)  & $|T_G|/|T_{OSN}|$ \\[0.5ex]  
\hline    
SNIB & $566,472$  & 2,001,333  & 7,273,177  &  $26\%$  \\  
DBLP & $900,440$  & 2,243,827  & 9,363,166  &  $25\%$   \\ 
\hline    
\end{tabular}
\label{table:dataset}
\end{table}

\begin{figure}
  \centering
  \subfigure[Data loading time. ]{
    \label{subfig:loadtime}
    \includegraphics[width=1.4in]{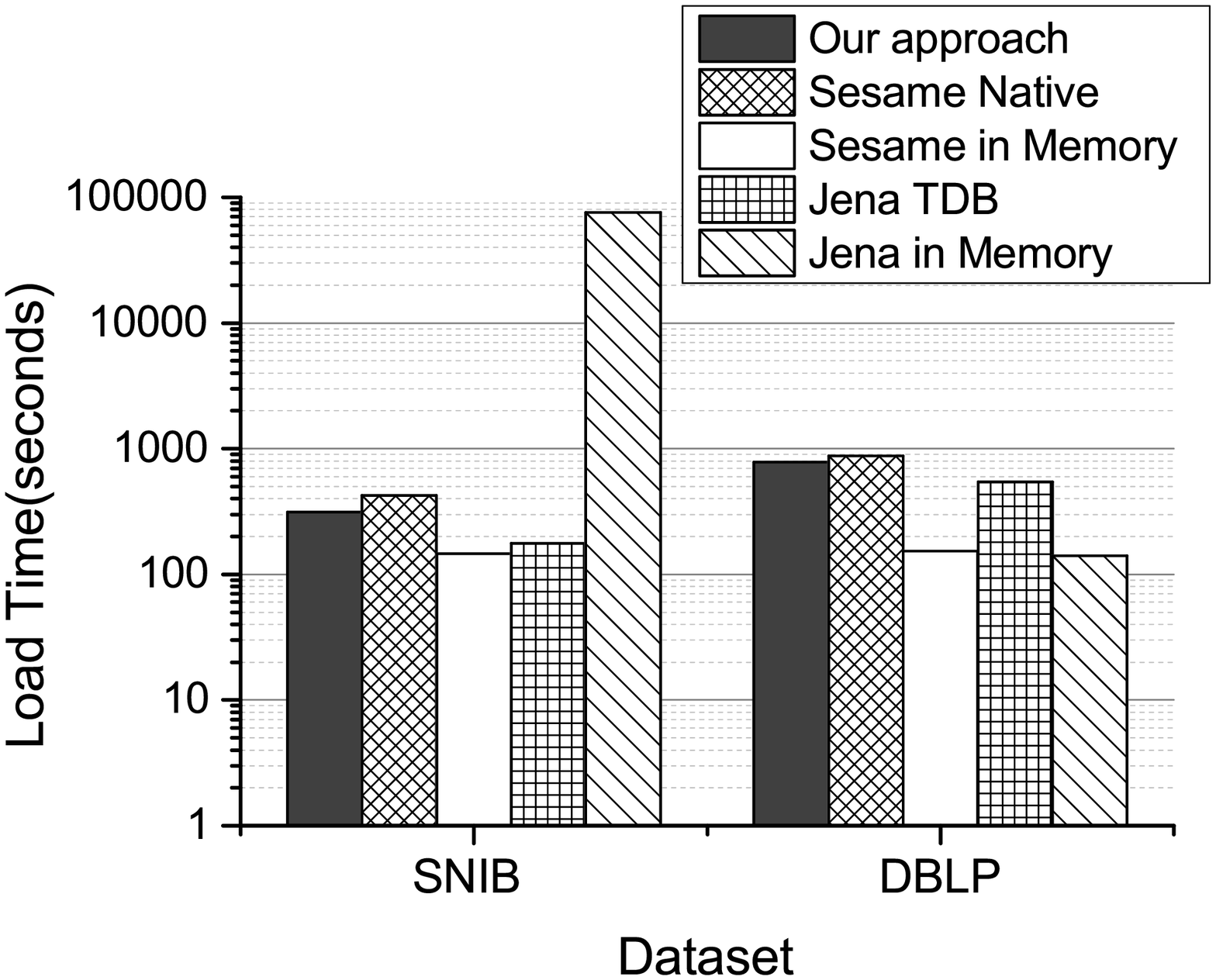}}
  \subfigure[Disk usage. ]{
    \label{subfig:diskusage}
    \includegraphics[width=1.4in]{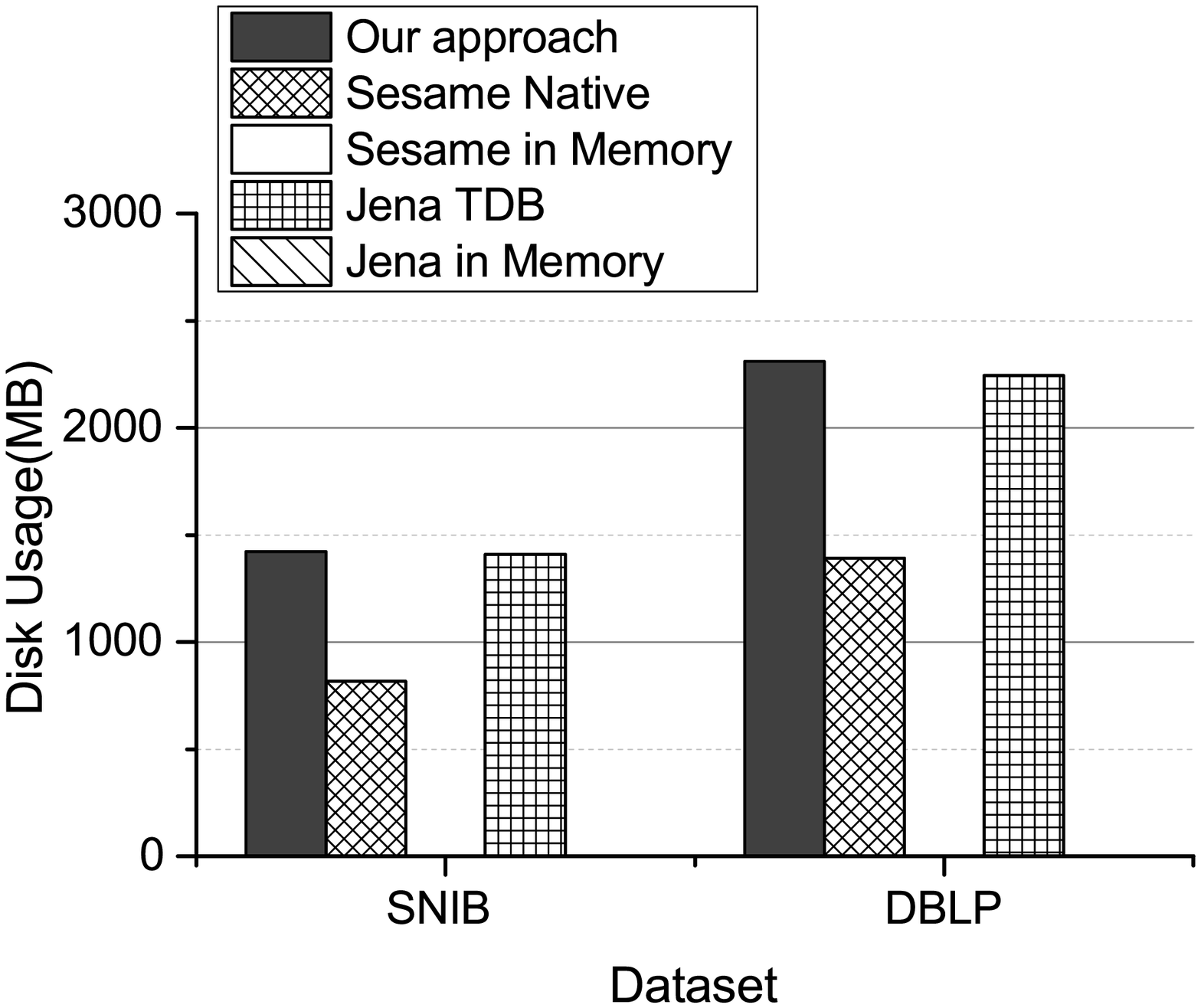}}
  \subfigure[Memory usage. ]{
    \label{subfig:memusage}
    \includegraphics[width=1.4in]{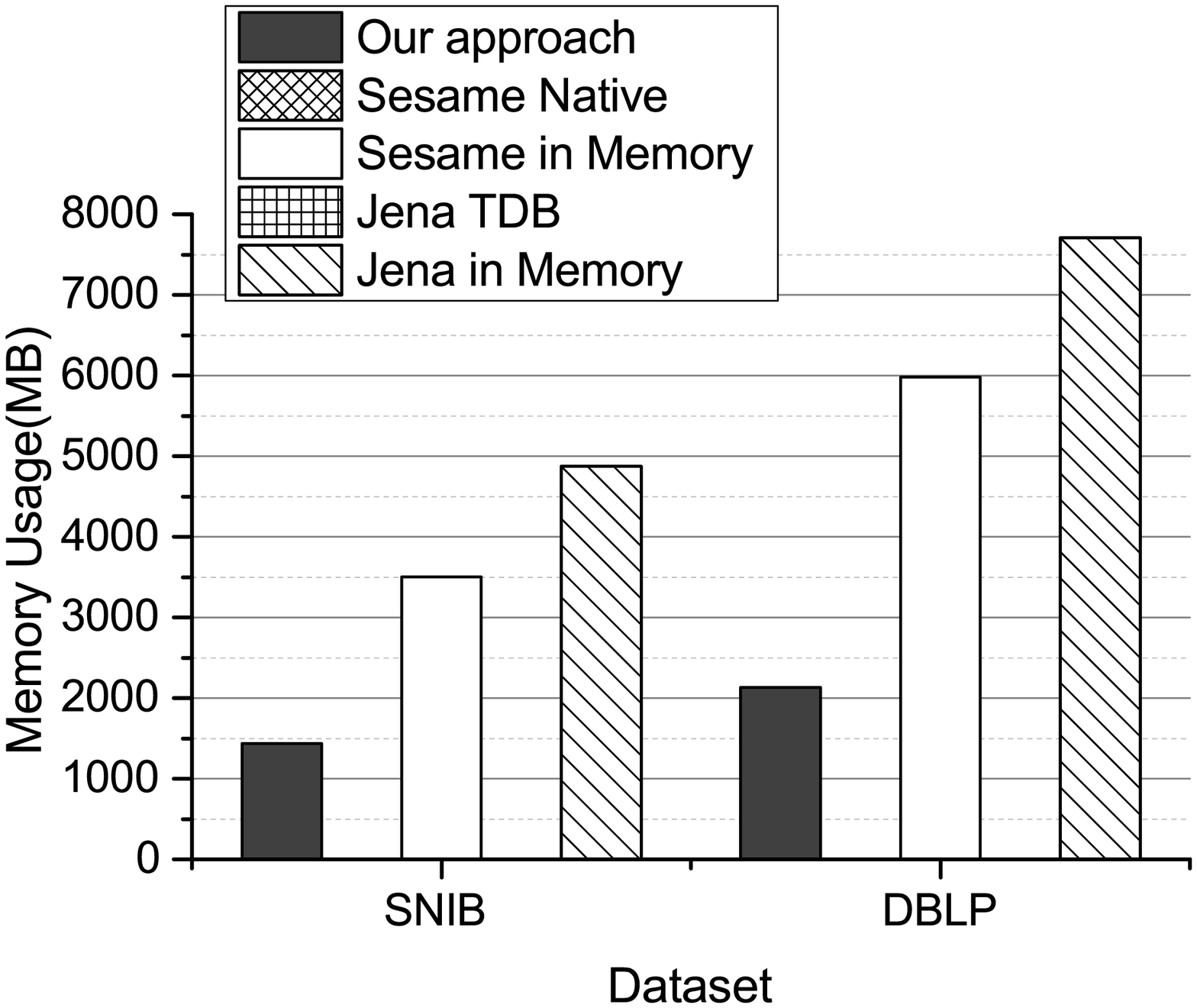}}
  \caption{\textit{Offline performance evaluation.} }
  \label{fig:offline} 
\end{figure}

For offline data loading time and memory expenses, we compare our approach with four competitors, Sesame and Jena in-memory store which store and index data only in memory, Sesame native store and Jena TDB which use disk as triple storage. Results of data loading time is represented in Figure \ref{subfig:loadtime},and disk usage in Figure \ref{subfig:diskusage}, memory usage in Figure\ref{subfig:memusage}. For our approach only load graph topological data into main memory, it need fewer memory than that of Jena and Sesame in-memory store, but with a little overhead of the data loading time.

\begin{figure}
  \centering
  \subfigure[\textit{SNIB query.}]{
    \label{subfig:snib} 
    \includegraphics[height=1.5in, width=1.2 in]{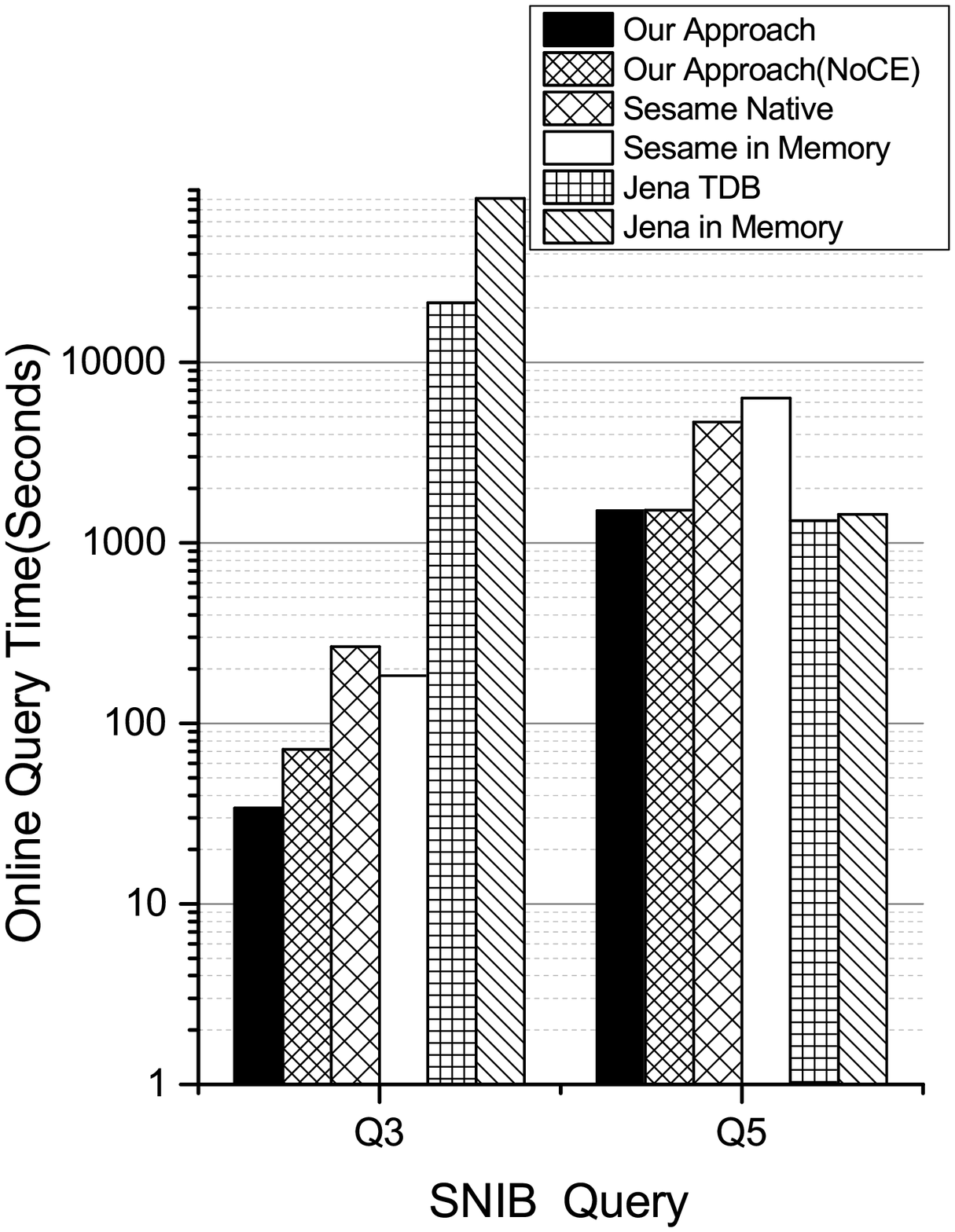}}
  \subfigure[\textit{DBLP query performance.}]{
    \label{subfig:gsparql} 
    \includegraphics[height=1.5in,width=2.8 in]{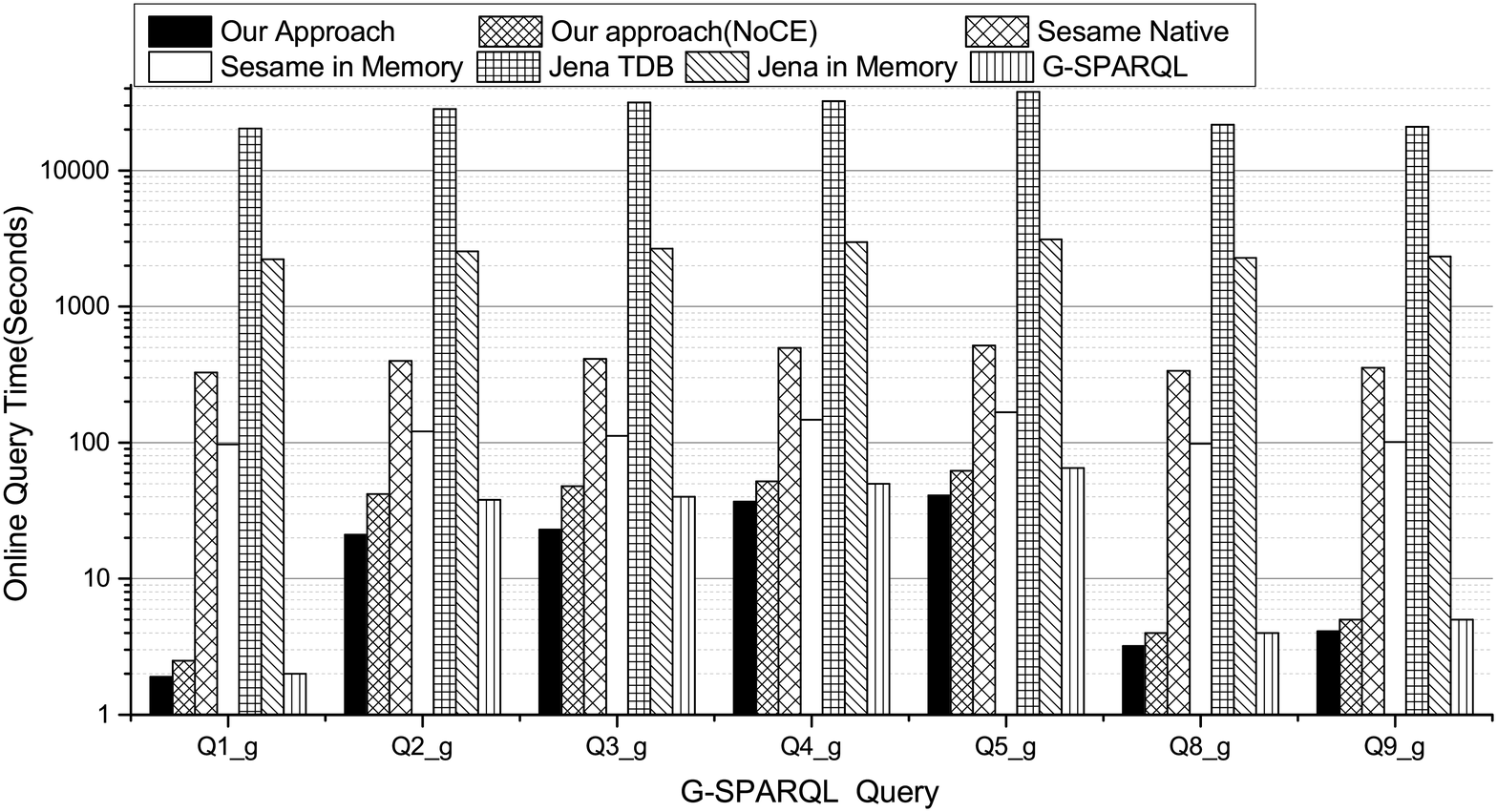}}
  \caption{\textit{Online Query Performance.}  }
  \label{fig:online} 
\end{figure}

For online query performance, in SNIB benchmark only $Q3$ and $Q5$ are path query related, while for DBLP we chosen 7 out of total 12 queries in G-SPARQL experiments(denoted as $Q_\_g$). In order for comparison, we had to rewrite these queries on DBLP dataset. Figure \ref{subfig:snib} shows that our approach achieved the best performance for SNIB $Q3$. As for SNIB $Q5$, the 3-HOP which expressed in UNION clause is explicitly parsed into six joins. This causes an expensive join expenses. Results in Figure \ref{subfig:gsparql} show that our approach works better than G-SPARQL(got directly from the \textit{Large Graph Size} result in \cite{sakr2012g}), while \textit{NoCE} has approximately the same performance as that of G-SPARQL. This shows that cost estimation for optimal join order can enhance the overall query performance.
\section{Related Works}
Most existing RDF stores uses a relational model to manage data, either in a traditional RDBMS or using a native triple store. They all processes SPARQL queries as sets of join operations using disk-based indices, which are costly for sophisticated joins. Some researches have focused on compressing and managing RDFs in main memory, Trinity RDF\cite{zeng2013distributed} is the most prominent among them. It uses graph exploration instead of join operations and greatly boosts SPARQL query performance. But manages all data in memory is not trivial, distributed shared memory increases the complexity for maintenance.

From graph analysis perspective, Property path can be viewed as the label-constraint reachability problem on labeled graph. Though reachability on graph is a further investigated problem, few literatures\cite{fan2011adding, zou2014efficient} considered its usage in property graph, which is more common in nature. These researches are mainly focus on building reachability indices in advance, which is time or space consuming and not adequate for large-scale data management.

Besides Jena and Sesame, several frameworks and prototypes have been proposed for path queries\cite{janik2005brahms, sakr2012g, gubichev2013sparqling}. BRAHMS\cite{janik2005brahms} only supports query on paths with predefined length. Sparqling Kleene\cite{gubichev2013sparqling} realizes join based on pre-constructed reachability indices which are space consuming. Our work is mainly motivated by G-SPARQL\cite{sakr2012g} which use the same hybrid storage and manage graph topological data in memory. Our work different from G-SPARQL in that, first, we are not design a new query language but uses standard SPARQL 1.1 instead, this makes our work more general. Second, G-SPARQL uses index-free pointer-based data structure to representing the graph topological data in memory, in order to suit for most graph algorithms. Our work only cares about path patterns which BFS algorithm is used to answer such reachability queries. We build simple indices only for facilitating BFS.
\section{Conclusion}\label{sec:conclusion}
In this paper we addressed the problem of property path query in RDF data, presented a step towards
incorporation of in-memory storage. In our approach we are not trying to invent new wheels, but managed to combine existing effective approaches as well as some technical enhancements. Contrast to traditional RDF management and graph query method, we used in-memory graph traversal instead of costly join to realize path query operator, used simple graph indices other than RDF permutation indices and complex graph reachability indices for efficient graph traversal. Evaluations have shown that our approach is feasible and efficient for process SPARQL property path queries.


\end{document}